\documentclass[aoas,nameyear,seceqn,dvips]{arximspdf}
\usepackage{graphics}
\doi{10.1214/09-AOAS297}
\volume{4}
\issue{2}
\pubyear{2010}
\firstpage{871}
\lastpage{892}

\makeatletter
\newcommand{\eqref}[1]{(\ref{#1})}
\renewcommand{\citet}[1]{[\citeauthor{#1} (\citeyear{#1})]}
\renewcommand{\citep}[1]{\citeauthor{#1} (\citeyear{#1})}
\makeatother

\begin{document}
\begin{frontmatter}

\title{Statistical inference of transmission fidelity of DNA
methylation patterns over somatic cell divisions in mammals}
\runtitle{Transmission fidelity of DNA methylation}

\begin{aug}
\author[a]{\fnms{Audrey Qiuyan} \snm{Fu}\corref{}\ead[label=e1]{audrey@stat.washington.edu}},\thanksref{m1,t3}
\author[b]{\fnms{Diane P.} \snm{Genereux}\ead[label=e2]{genereux@u.washington.edu}},\thanksref{t2,t3}
\author[b]{\fnms{Reinhard} \snm{St\"{o}ger}\ead[label=e4]{stoeger@u.washington.edu}\ead[label=e7]{Reinhard.Stoger@Nottingham.ac.uk}},\thanksref{m2,t3}
\author[b]{\fnms{Charles D.} \snm{Laird}\ead[label=e5]{cdlaird@u.washington.edu}}\thanksref{t3}
\and
\author[c]{\fnms{Matthew} \snm{Stephens}\ead[label=e6]{mstephens@uchicago.edu}}

\thankstext{m1}{Current address: Cambridge Systems Biology Centre,
University of Cambridge,
Tennis Court Road,
Cambridge, CB2 1QR, UK.}
\thankstext{t2}{Supported in part by NIH Training Grant T32 HG00035 to
the University of Washington.}
\thankstext{m2}{Current address:
University of Nottingham,
School of Biosciences,
Division of Animal Sciences,
Sutton Bonington Campus,
Loughborough, Leicester, LE12 5RD, UK.
\printead{e7}.}
\thankstext{t3}{Supported in part by NIH Grants HD002274 and GM077464.}
\runauthor{A. Q. Fu et al.}
\affiliation{University of Washington, University of Washington,
University of Washington, University of Washington and University of Chicago}
\address[a]{A. Q. Fu\\
Department of Statistics\\
University of Washington\\
Seattle, Washington 98195\\
USA\\ \printead{e1}}
\address[b]{D. P. Genereux\\
R. St\"{o}ger\\
C. D. Laird\\
Department of Biology\\
University of Washington\\
Seattle, Washington 98195\\
USA\\
E-mails: \printead*{e2}\\
\phantom{E-mails: }\printead*{e4}\\
\phantom{E-mails: }\printead*{e5}}
\address[c]{M. Stephens\\
Departments of Human Genetics\\
\quad and Statistics\\
University of Chicago\\
Chicago, Illinois\\
USA\\
\printead{e6}}
\end{aug}

\received{\smonth{6} \syear{2009}}
\revised{\smonth{9} \syear{2009}}

\begin{abstract}
We develop Bayesian inference methods for a recently-emerging type of epigenetic data to study
the transmission fidelity of DNA methylation patterns over cell divisions.
The data consist of parent-daughter double-stranded DNA methylation patterns with each pattern coming
from a single cell  and represented as an unordered pair of binary strings.
The data are technically difficult and time-consuming to collect, putting a premium on an
efficient inference method.  Our aim is to estimate rates for the maintenance
and de novo methylation events that gave rise to the observed patterns, while accounting for measurement error.
We model data at multiple sites jointly,
thus using whole-strand information, and considerably reduce confounding between parameters.
We also adopt a hierarchical structure that allows for variation in rates across sites
without an explosion in the effective number of parameters.
Our context-specific priors capture the expected stationarity, or near-stationarity, of the
stochastic process that generated the data analyzed here.  This expected stationarity is shown to greatly
increase the precision of the estimation. Applying our model to a data set collected at the human \textit{FMR1}
locus, we find that
measurement errors, generally ignored in similar studies,
occur at a nontrivial rate (inappropriate bisulfite conversion error: 1.6$\%$ with 80$\%$ CI: 0.9--2.3$\%$).
Accounting for
these errors has a substantial impact on estimates of key biological parameters.
The estimated average failure of maintenance rate and daughter de novo rate decline from 0.04 to 0.024
and from 0.14 to 0.07, respectively, when errors are accounted for.  Our results also provide evidence that de novo
events may occur on both parent and daughter strands: the median parent and daughter
de novo rates are 0.08\break (80$\%$ CI: 0.04--0.13) and 0.07 (80$\%$ CI: 0.04--0.11), respectively.
\end{abstract}

%

\end{frontmatter}
%

\section{Introduction}
\label{sec:introduction}
In this paper we develop statistical models
and inference methods to address an important problem in epigenetic biology:
inference of the fidelity with which DNA methylation patterns
in DNA are preserved over somatic cell divisions in mammals.
The double-stranded DNA methylation data we present here have the
potential to yield important biological insights.
However, due to limitations of current experimental technologies, these
data also present challenges. For example, it is difficult
to obtain this type of data in large quantities, some key biological
variables are unobservable, certain parameters of interest
may be confounded, and the data are subject to
measurement error at perhaps a nontrivial rate \citet{Genereuxetal2008}. These characteristics put a premium on efficient
inference methods that make
full use of the data while dealing with complexities intrinsic to the
data and to the biological problem. In this section
we introduce, for a statistical audience, relevant biological
background on DNA methylation and the hairpin-bisulfite PCR technique
[\citep{Lairdetal2004}; \citep{Mineretal2004}]
used to collect the data. We then state our aim and give overviews of
existing methods and our new approach.

A DNA molecule is most commonly described by its sequence of nucleotides,
consisting of adenine, cytosine, guanine and thymine; or A, C, G and T.
However, this description is incomplete in that it omits some
functionally relevant
features. An important example is that some nucleotides are
\textit{methylated}---that is, at some nucleotide positions a methyl group
has been chemically attached to the DNA. Methylation is an epigenetic
mechanism, such that the pattern of methylation along
a DNA molecule can profoundly effect its function. Aberrant methylation
plays a role in many cancers [\citep{ChenRiggs2005};
\citep{JonesBaylin2002}; \citep{Laird2003}] and in several human developmental diseases, including
fragile X
syndrome [\citep{Laird1987}; \citep{RobertsonWolffe2000}]. In fragile X syndrome,
hypermethylation
of the \textit{FMR1} locus on the X chromosome leads to many
manifestations, including
mental retardation. A critical distinction between methylation patterns
and nucleotide sequences is that,
whereas the latter are generally assumed to be identical in nearly all
cells of an organism, the former can vary considerably
from cell to cell \textit{within} an organism.The processes that govern
DNA methylation and its variability across cells
are thus of considerable biological interest.

In mammals, methylation of DNA occurs almost exclusively on cytosines
(Cs) that are followed by
a guanine (G), locations referred to as CpG sites.
On average, about 70--80\% of CpG sites in mammals are methylated
\mbox{\citet{Ehrlichetal1982}}.
A key property of DNA molecules is that they are double-stranded, with
the two strands
complementary to each other, that is, A pairs with T, and C with G.
Hence, we also refer
to CpG sites as CpG/CpG dyads to emphasize their double-strandedness.
At a CpG/CpG dyad,
methyl groups can be present on both strands (which we call
``methylated''), on one strand
(``hemimethylated''), or on neither strand (``unmethylated''). Our
focus here is on the accuracy with which the pattern of methylation on
one strand
(the parent strand) of a DNA molecule is transferred to the new
complementary strand
(the daughter strand) produced by DNA replication [see Supplementary
Material Section 1
in \citep{Fuetal2009supp} for details].

\begin{figure}[b]

\includegraphics{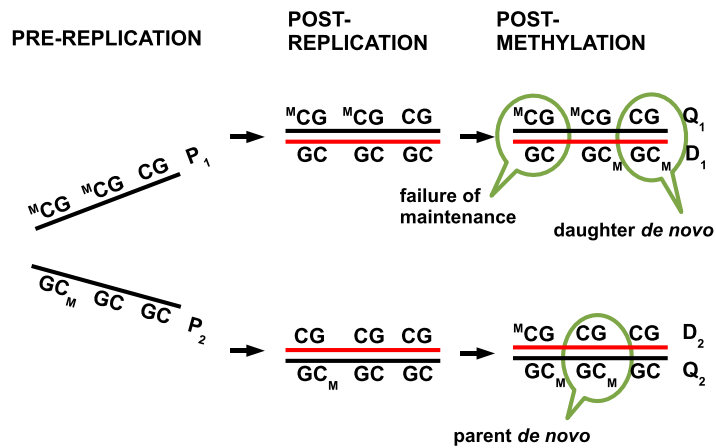}

\caption{The transmission process of DNA methylation patterns in
mammalian somatic
cells. The two strands in a DNA molecule each become parent strands
during DNA
replication (from left column to middle column), used as a template to
synthesize
a daughter strand (red lines). During the
short, intermediate stage (middle column), daughter strands
are completely unmethylated, whereas parent strands have the same
methylation patterns
as before replication. Subsequently, methyl groups are added to
cytosines (right column).
Failure of maintenance and de novo methylation events can occur,
leading to differences in
methylation patterns on parent and daughter strands. Binary vectors~$\mathbf{P}
_i$,~$\mathbf{Q}_i$ and
$\mathbf{D}_i$, where $i=1,2$, denote methylation patterns on a pre-replication
parent strand, on a post-replication parent strand and
on a daughter strand, respectively.}
\label{fig:trans.proc}
\end{figure}

The transmission process of methylation patterns is complex and
imperfect: cytosines are first
incorporated into DNA and subsequently methylated. Sometimes, however,
a cytosine on the daughter strand
remains unmethylated even when the parent is methylated, an event we
refer to
as a \textit{failure of maintenance} event. Methylation is also sometimes
introduced at
previously unmethylated locations; we refer to such events as \textit{de
novo} methylation events.
Here, as in \citep{Genereuxetal2005}, we allow the possibility of de
novo events on both parent and daughter strands.
Figure \ref{fig:trans.proc} illustrates these concepts
under a widely-accepted model for the transmission process [Bird (\citeyear{Bird2002,Bird2007})].

Here we consider the problem of using double-stranded DNA methylation patterns
to estimate the rates at which failure of maintenance and de novo
methylation events occur.
We collected these double-stranded data using hairpin-bisulfite PCR
\citet{Lairdetal2004},
which was modified as in \citep{Mineretal2004} to include molecular
codes to authenticate each DNA
methylation pattern, removing redundant and contaminant patterns
[details of the experimental design
are in Supplementary Material Section 2 in \citep{Fuetal2009supp}].
Several features of hairpin-bisulfite PCR are particularly relevant to
statistical modeling:
(i) a short ``hairpin'' DNA sequence links together complementary parent
and daughter strands;
(ii) linked strand pairs are subject to bisulfite conversion which
reveals their double-stranded
methylation patterns; and (iii) errors arise due to imperfections in
the bisulfite conversion process
\citet{Genereuxetal2008}. Thus, this method yields data, subject to
measurement error due to bisulfite
conversion, on methylation patterns for parent-daughter
pairs from individual molecules. Current experimental technologies,
however, cannot determine strand type; that is, which strand is the
parent and which the daughter.

Data on double-stranded methylation patterns obtained by
hairpin-bisulfite PCR have previously been analyzed by \citep
{Lairdetal2004} and
\citep{Genereuxetal2005}. The analysis in \citep{Lairdetal2004},
which is not explicitly likelihood-based, involves
counting the number of events of each type at each site over strands
and then averaging the counts over the two possible assignments of
strand identity,
assuming that de novo events occur only on the daughter strand. \mbox{\citep{Genereuxetal2005}} assumed strict
stationarity of the stochastic process that generates the data and
based their analysis
on a likelihood function for individual CpG sites without incorporating
information about which observations
at different sites in a double-stranded molecule are on the same
strand, and which are on different strands.
These existing analyses provide the foundations for our work here.

Here we develop a
full statistical model for the data, exploiting information from
contiguous sites rather than from individual sites alone.
Three additional innovations of our
modeling approach are as follows: (i) accounting for measurement
errors, which are due to imperfections in the bisulfite conversion process;
(ii) relaxing the strict stationarity assumption made in \citep
{Genereuxetal2005}; and
(iii)~using a hierarchical structure to allow rates of key parameters
to vary across sites without greatly increasing the effective size of
the parameter space.

\section{Models and methods}
\label{sec:methods}
\subsection{Basic model and key assumptions}
\label{sec:methods.basic}
We consider data collected using hairpin-bisulfite PCR,
on methylation states at $S$ CpG sites on $N$ double-stranded DNA molecules.
We denote an unmethylated CpG site by 0, and a methylated CpG site by
1, so
the data are $N$ pairs of binary strings, $\{\mathbf{x}_1,\mathbf
{y}_1\},\ldots,\{\mathbf{x}_N,\mathbf{y}
_N\}$, each string being of length $S$.
Current technologies are not able to identify strand type; that is, we
do not know which data vector ($\mathbf{x}_i$ or $\mathbf{y}_i$)
arose from the parent strand
and which from the daughter. Hence, we use $\{ \cdot\}$ to represent this
lack of ordering in each observed pair.
We initially assume that the data are observed without error and then
relax this assumption.

Our model introduces latent random variables $\mathbf{Q}_i$ and
$\mathbf{D}_i$, each a
binary vector, representing
patterns of methylation on the parent strand and daughter strand, respectively.
These binary vectors may be thought of as potentially-imperfect copies of
the patterns of methylation on the unobserved pre-replication parent
strand, which we denote by binary vector $\mathbf{P}_i$ (Figure \ref
{fig:trans.proc}). Differences between $\mathbf{P}_i$ and $\mathbf
{D}_i$ can arise
due to failure of maintenance, or de novo methylation on the daughter
strand; differences between $\mathbf{P}_i$ and $\mathbf{Q}_i$ can
arise due to de novo
methylation
on the parent strand. We assume that these three types of events occur
independently of one another, and independently across individuals and
across sites.
Denoting the probabilities of these events at site $j$ by $1-\mu_j$,
$ {\delta_d}_j$ and $ {\delta_p}_j$, respectively, and assuming no
spontaneous loss
of methylation
on the parent strand (explained below), we have
%
\begin{eqnarray}
\qquad \Pr(D_{ij} = 0 | P_{ij} = 1) &=& 1-\mu_j  \qquad\mbox{(failure of
maintenance)}, \label{eqn:failure.of.maint}\\
\Pr(Q_{ij} = 1 | P_{ij} = 0) & =& {\delta_p}_j \qquad \mbox{(de novo
methylation
on parent)}, \label{eqn:parent.denovo}\\
\Pr(D_{ij} = 1 | P_{ij} = 0) & =& {\delta_d}_j \qquad \mbox{(de novo
methylation
on daughter)}. \label{eqn:daughter.denovo}
\end{eqnarray}
We are interested in estimating failure of maintenance and de novo methylation
rates at CpG sites and assessing their variability across sites. We use
$\lambda=\{\mu_j, {\delta_p}_j, {\delta_d}_j, j=1,\ldots,S\}$ to
denote the vector of
parameters.

To derive the likelihood function for those parameters, we make three
key assumptions.
The first assumption is that there is no active removal of methyl
groups on
the parent strand. That is, if the parent strand is
methylated before replication, then it will also be methylated after
replication:
%
\begin{equation}
\Pr(Q_{ij}=1 | P_{ij}=1) = 1. \label{eqn:no.demeth}
\end{equation}
Although recent publications, such as \mbox{\citep{Metivieretal2008}}
and\break
\citep{Kangaspeskaetal2008}, suggest the
possibility that transcriptionally \textit{active} loci
can have very rapid changes in methylation patterns which may be due to active
removal of methyl groups from the template DNA, there is no evidence so
far that this
active removal occurs at \textit{inactive} loci in leukocytes, the locus
type and the cell type from which our
data were collected. This assumption is also consistent with what
underlies the models in \citep{Lairdetal2004}
and \citep{Genereuxetal2005}. The conditional probability in (\ref
{eqn:no.demeth}) then joins
those in (\ref{eqn:failure.of.maint})--(\ref{eqn:daughter.denovo}) to
form a complete
probabilistic characterization of the transmission process at a single CpG
site.

The second assumption is that methylation events occur independently of
one another across
sites. Equations (\ref{eqn:failure.of.maint})--(\ref{eqn:no.demeth}),
together with this
assumption, determine
the conditional distribution of the \textit{ordered} pair $(\mathbf
{Q}_i,\mathbf{D}_i)$
given $\mathbf{P}_i$, which we denote $h_\lambda$ (Table \ref
{table:trans_prob}).
To complete the specification of the distribution of $(Q_{ij},
D_{ij})$, we further model $P_{ij}$s
as independent Bernoulli random variables with methylation probability $m_j$:
%
\begin{equation}
\operatorname{Pr} (P_{ij} = 1) = m_j.
\end{equation}
Under this second assumption, we obtain the likelihood function for a
single double-stranded
methylation pattern with known strand type as the product of
probabilities of methylation patterns at
individual sites, each probability summing over two possibilities of
the methylation status
(represented by $p_{ij}$) on
the pre-replication parent strand $\mathbf{P}_i$. Specifically, we
give the
likelihood for the case where~$\mathbf{x}_i$ contains data from the parent strand and $\mathbf
{y}_i$ contains data
from the daughter strand:
%
\begin{eqnarray}\label{eqn:lik.simple.d}
&&\Pr\bigl((\mathbf{Q}_i,\mathbf{D}_i)=(\mathbf{x}_i, \mathbf{y}_i)|
\lambda\bigr)
\nonumber
\\[-8pt]
\\[-8pt]
\nonumber
&&\qquad =  \prod^S_{j = 1} \sum
^1_{p_{ij} = 0} h_\lambda(x_{ij}, y_{ij}; p_{ij}) m_j^{p_{ij}}
(1-m_j)^{1-p_{ij}}.
\end{eqnarray}

%
\begin{table}
\tabcolsep=0pt
\caption{Probabilities of methylation events at site $j$,
$h_\lambda(q_{ij},d_{ij};p_{ij}) =\Pr( (Q_{ij},D_{ij}) =
(q_{ij},d_{ij}) | P_{ij} = p_{ij} )$}
\label{table:trans_prob}
\begin{tabular*}{\textwidth}{@{\extracolsep{\fill}}cccl@{}}
\hline
\multicolumn{1}{@{}l}{$\bolds{(Q_{ij}, D_{ij}) = (q_{ij},d_{ij})}$} & $\bolds{P_{ij} = p_{ij}}$ & $\bolds{h_\lambda
(q_{ij},d_{ij};p_{ij})}$ & \multicolumn{1}{c}{\textbf{Methylation event}} \\
\hline
(0, 0) & 1 & 0 & $\dagger$Assumed not to occur \\
(0, 1) & 1 & 0 & $\dagger$Assumed not to occur \\
(1, 0) & 1 & $1 - \mu_j$ & Failure of maintenance \\
(1, 1) & 1 & $\mu_j$ & Maintenance \\
(0, 0) & 0 & $(1 - \delta_{p j}) (1 - \delta_{d j})$ & No de novo on
parent or daughter\\
(0, 1) & 0 & $(1 - \delta_{p j}) \delta_{d j}$ & De novo on daughter
but not parent \\
(1, 0) & 0 & $\delta_{p j} (1 - \delta_{d j})$ & De novo on parent
but not daughter \\
(1, 1) & 0 & $\delta_{p j} \delta_{d j}$ & De novo on parent and
daughter \\
\hline
\end{tabular*}
\tabnotetext[]{tz}{The dagger $\dagger$ indicates cases not possible
under the assumption of no active removal of methyl groups on the
parent strand.}
\end{table}

Since strand type is unobserved, to get the probability of the observed
double-stranded methylation
pattern $i$, we must sum over the two possible assignments of strand type:
%
\begin{eqnarray}
\hspace*{20pt}&&\Pr( \{\mathbf{Q}_i,\mathbf{D}_i\} = \{\mathbf{x}_i,\mathbf{y}_i\}
| \lambda)
\nonumber
\\[-8pt]
\\[-8pt]
\nonumber
&&\qquad= \biggl(\frac{1}{2}
\biggr)^{\mathbf{1}(\mathbf{x}_i=\mathbf{y}_i)} \bigl\{\Pr\bigl((\mathbf
{Q}_i, \mathbf{D}_i) = (\mathbf{x}_i, \mathbf{y}_i) |
\lambda\bigr)+ \Pr\bigl((\mathbf{Q}_i, \mathbf{D}_i) = (\mathbf{y}_i, \mathbf
{x}_i) | \lambda\bigr) \bigr\},\hspace*{-20pt}
\end{eqnarray}
where $\mathbf{1}(A)$ is the indicator function, taking value 1 if
condition $A$ is true and 0 otherwise.

By making the third assumption that data from the $N$ double-stranded
methylation patterns are independent draws
from the same distribution with parameter $\lambda$, we then obtain a
likelihood function of $\lambda$ for all $N$
patterns:
%
\begin{eqnarray}
\hspace*{15pt}L (\lambda; \{\mathbf{x},\mathbf{y}\}) &=& \prod^N_{i = 1} \Pr( \{
\mathbf{Q}_i,\mathbf{D}_i\} = \{\mathbf{x}_i,\mathbf{y}
_i\}| \lambda) \\
& \propto&\prod_{i=1}^N \bigl\{\Pr\bigl((\mathbf{Q}_i, \mathbf{D}_i) =
(\mathbf{x}_i, \mathbf{y}_i)| \lambda
\bigr)+ \Pr\bigl((\mathbf{Q}_i, \mathbf{D}_i) = (\mathbf{y}_i, \mathbf
{x}_i)| \lambda\bigr) \bigr\}.\hspace*{-26pt} \label{eqn:lik.simple}
\end{eqnarray}

\subsection{Incorporating measurement error and estimating error rates}
\label{sec:models.error}

As mentioned in Section \ref{sec:introduction}, imperfection in
bisulfite conversion is an important source of potential measurement
error here and in other applications involving bisulfite conversion. In
brief, bisulfite conversion is an experimental technique that
aims to convert unmethylated cytosines to a different base, uracil,
thus allowing unmethylated
and methylated locations to be identified by DNA sequencing.
Imperfections during this process can lead to two types of error:
failure of
conversion, where bisulfite fails to convert an unmethylated cytosine
(resulting in a truly unmethylated site being measured as methylated)
and inappropriate conversion, where bisulfite converts a methylated
cytosine to a thymine (leading to a truly
methylated site being measured as unmethylated). We let $b=(b_1,\ldots
,b_S)$ and $c=(c_1,\ldots,c_S)$ denote the respective rates at which
these two types of errors occur, where the elements~$b_j$ and $c_j$
represent the error rates at site $j$.

To incorporate these measurement errors into the model, we
introduce random variables $Q'_{ij}$ and $D'_{ij}$ to denote
the \textit{observed} methylation states on the post-replication parent
strand and
the daughter strand, while continuing to use $Q_{ij}$ and~$D_{ij}$
to denote \textit{true} methylation states on those two strands.
We assume that errors occur independently across CpG sites and DNA
strands, so that
the conditional distribution of the observed data given the true states
is given by
%
\begin{eqnarray}
&&\Pr\bigl((Q'_{ij},D'_{ij})=(x_{ij},y_{ij})|(Q_{ij},D_{ij})\bigr)
\nonumber
\\[-8pt]
\\[-8pt]
\nonumber
&&\qquad=\Pr
(Q'_{ij}=x_{ij}|Q_{ij})
\Pr(D'_{ij}=y_{ij}|D_{ij}),
\end{eqnarray}
where each term on the right-hand side is a function of bisulfite
conversion error rates $b_j$ and $c_j$ as
in Table \ref{table:error.defs}.

We extend the parameter vector $\lambda$ to incorporate these
measurement error parameters,
$\lambda=\{\mu, {\delta_p}, {\delta_d}, b, c\}$. The
likelihood function, allowing for measurement error, becomes
%
\begin{eqnarray} \label{eqn:lik.error}
&&L (\lambda; \{ \mathbf{x}, \mathbf{y}\})
\nonumber
\\[-8pt]
\\[-8pt]
\nonumber
&&\qquad\propto \prod^N_{i=1} \bigl\{ \Pr\bigl((\mathbf{Q}'_i, \mathbf{D}'_i)
= (\mathbf{x}_i, \mathbf{y}_i)| \lambda
\bigr)+ \Pr\bigl((\mathbf{Q}'_i, \mathbf{D}'_i) = (\mathbf{y}_i, \mathbf
{x}_i)| \lambda\bigr) \bigr\},
\end{eqnarray}
where
%
\begin{eqnarray}\label{eqn:lik.error.d}
&&\qquad \Pr\bigl((\mathbf{Q}'_i, \mathbf{D}'_i) = (\mathbf{x}_i, \mathbf{y}_i)|
\lambda\bigr)\nonumber\\
&&\qquad\qquad= \prod^S_{j = 1} \sum
^1_{q_{ij}=0} \sum^1_{d_{ij}=0}
\Pr\bigl((Q'_{ij}, D'_{ij}) = (x_{ij}, y_{ij}) | (Q_{ij}, D_{ij}) =
(q_{ij},d_{ij})\bigr)\\
&&\qquad\qquad{}\hspace*{70pt}\times\sum^1_{p_{ij} = 0} h_{\lambda}(q_{ij}, d_{ij}; p_{ij})
m_j^{p_{ij}} (1 - m_j)^{1 - p_{ij}},\nonumber
\end{eqnarray}
%
and $\Pr((\mathbf{Q}'_i, \mathbf{D}'_i) = (\mathbf{y}_i, \mathbf
{x}_i)| \lambda)$ is defined similarly.

%
\begin{table}[t]
\tablewidth=267pt
\caption{Rates of bisulfite conversion error, which are conditional
probabilities of the observed methylation state being different from a~given true methylation
state. Specifically, $b_j$ is the failure of conversion rate at site
$j$ and $c_j$ the inappropriate conversion rate}
\label{table:error.defs}
\begin{tabular*}{267pt}{@{\extracolsep{4in minus 4in}}lc@{\hspace*{6pt}}cc@{}}
\hline
& & \multicolumn{2}{c}{\textbf{Observed} \textbf{(}$\bolds{Q'_{ij}}$ \textbf{or} $\bolds{D'_{ij}}$\textbf{)}}
\\[-6pt]
& & \multicolumn{2}{c@{}}{\hrulefill}\\
& & \textbf{0} & \textbf{1} \\
\hline
\textbf{Truth ($\bolds{Q_{ij}}$ or $\bolds{D_{ij}}$)} & \textbf{0} & $1-b_j$ & $b_j$ \\
& \textbf{1} & $c_j$ & $1-c_j$ \\
\hline
\end{tabular*}
\end{table}
%


\subsection{Hierarchical model for variability in rates across sites}

In the above formulation we have allowed that rates may take different
values across sites $j=1,\ldots,S$.
In practice, there is limited information about the rates at any given
site, so attempting to estimate
each of these parameters separately will produce highly variable estimates.
To overcome this challenge, we employ a hierarchical model to effectively
reduce the dimensionality of the parameter space and to borrow strength
across sites.
In this hierarchical model we assume that the components of the vectors
$\mu$, ${\delta_p}$, ${\delta_d}$, $m$ and $c$ each follow a beta
distribution.

In specifying these beta distributions, we find it convenient to use
the parameterization
$\operatorname{Beta}(r,g)$ to denote the beta distribution with mean $r$ and variance
$g r(1-r)$, hence referring to the parameter
$g$ as the ``scaled'' variance. The relationship between this
parametrization and the conventional
$\alpha$--$\beta$ parametrization is
%
\begin{equation}
r=\frac{\alpha}{\alpha+\beta}, \qquad  g=\frac{1}{\alpha+\beta+1}
\label{eqn:beta.rg}
\end{equation}
for a $\operatorname{Beta}(\alpha,\beta)$ random variable $X$ with density
%
\begin{equation}
f(x) \propto x^{\alpha-1}(1-x)^{\beta-1}.
\end{equation}
We prefer the $r$--$g$ parameterization in our analysis because (i) $r$
and $g$ are easily interpretable;
and (ii) this parametrization facilitates specification of sensible
priors---in particular,
it is reasonable to assume $r$ and $g$ to be independent a priori.

Our hierarchical model assumes a separate set of $r$ and $g$ for each
of the vectors~$\mu$,
${\delta_p}$, ${\delta_d}$, $b$ and $c$:
%
\begin{eqnarray}
\mu_j & \sim&\operatorname{Beta}(r_\mu,g_\mu), \label{eqn:beta.mu}\\
{\delta_p}_j & \sim&\operatorname{Beta}(r_{dp},g_{dp}), \label{eqn:beta.dp}\\
{\delta_d}_j & \sim&\operatorname{Beta}(r_{dd},g_{dd}), \label{eqn:beta.dd}\\
b_j &\sim&\operatorname{Beta} (r_b, g_b), \label{eqn:beta.b}\\
c_j & \sim&\operatorname{Beta}(r_c,g_c). \label{eqn:beta.c}
\end{eqnarray}

The methylation probability vector, $m$, is dealt with slightly
differently, as described in the next section.


\subsection{Incorporating stationarity}

Previous analyses of these types of data \citet{Genereuxetal2005}
have been based on the assumption that the transmission
process has attained temporal stationarity; that is, at each site the
proportion of methylated CpGs
is stable over generations of cell division. Supporting
biological evidence for this assumption comes from observations
that methylation densities at the \textit{FMR1} locus were virtually
unchanged over a
five-year time span in several human males with fragile X syndrome
\citet{Stogeretal1997}.

The assumption of stationarity in \citep{Genereuxetal2005} imposes the
following strict relationship
between the methylation probability $m_j$ and the failure of
maintenance and de novo methylation rates:
%
%
\begin{equation}
\label{eqn:stat.m}
m_j = \frac{{\delta_p}_j+{\delta_d}_j}{1+{\delta_p}_j+{\delta
_d}_j-\mu_j}.
\end{equation}
Requiring strict equality in this equation appears to be a rather
strong assumption. Indeed, examples
in \citep{Fu2008} illustrate the strong effect this assumption can have
on the likelihood surface.

Thus, to avoid making this strong assumption, and to improve robustness
to departures from strict stationarity,
we exploit the flexibility of the Bayesian modeling approach to allow
for deviations
from strict equality in (\ref{eqn:stat.m}).

Specifically, to incorporate stationarity, we assume that each $m_j$
follows a beta distribution,
%
\begin{equation}
m_{j} \sim\operatorname{Beta} ({r_{m}}_{j}, g_{m}),
\end{equation}
with mean parameter
%
\begin{equation}
{r_{m}}_j = \frac{{\delta_p}_j + {\delta_d}_j}{1 + {\delta_p}_j +
{\delta_d}_j - \mu_j}.
\label{eqn:stat.rm}
\end{equation}
This distribution on $m_j$ is centered on its expected value under the
stationarity assumption,
but allows for deviations, measured by $g_m$: at a CpG site
small values of~$g_m$ represent near-stationarity, whereas large values
indicate substantial deviations from stationarity.

\subsection{Bayesian inference and choice of priors}

We choose to use a Bayesian approach to fit the hierarchical model,
specifying priors for the values of mean $r$ and scaled variance $g$ in
beta distributions (\ref{eqn:beta.mu})--(\ref{eqn:beta.c}).

We assign an independent uniform prior to each $r$: a $\operatorname{Uniform}(0,1)$
prior for each of
$r_\mu$, $r_{\delta_p}$ and $r_{\delta_d}$, and a $\operatorname{Uniform}(0,0.06)$
prior for $r_c$
because experimental results suggest that measurement error rate $c_j$
is likely to be below 0.06.
We can use a similar method to estimate $r_b$ (and $g_b$) for the other
error rate $b_j$, although in our data
analysis $b_j$ is fixed to an estimate obtained from experiments (see
Section~\ref{sec:results}).

\begin{figure}

\includegraphics{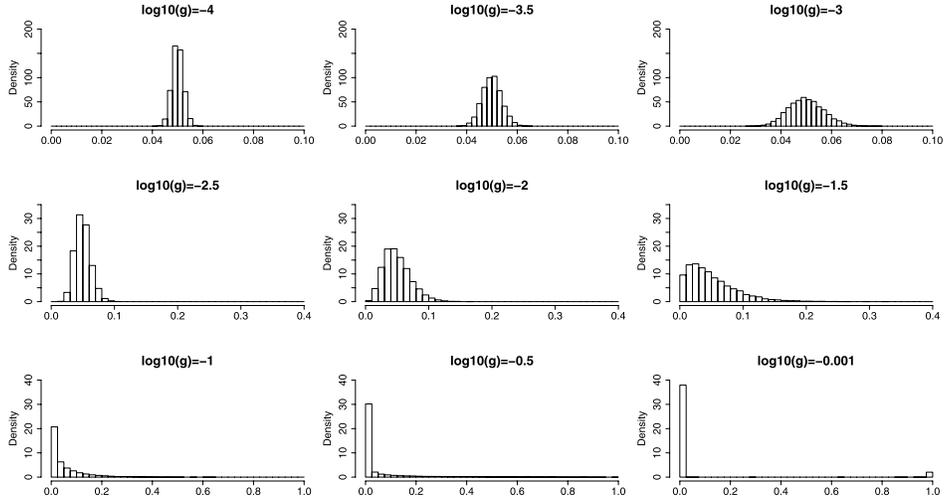}

\caption{Shape of a beta distribution changes with respect to scaled
variance $g$. Data were simulated for beta distributions
with mean $r=0.05$ and different values of scaled variance $g$. Ranges
of the horizontal and the vertical axes
are different between rows. As $g$ increases, the histogram spreads out
to the entire support
of $(0,1)$ and a second peak at 1 starts to appear (bottom right panel).}
\label{fig:beta-g}
\end{figure}

We assign a $\operatorname{Uniform}(-4,0)$ prior to each ${\log_{10}} g$. This choice of
prior has the flexibility of
capturing a wide range of beta distributions with qualitatively
different levels of variability.
Figure~\ref{fig:beta-g} illustrates this point: for a fixed mean value $r$,
as ${\log_{10}} g$ increases, the beta distribution becomes more and more
spread out over the support
$(0,1)$. In other words, this choice of prior on ${\log_{10}} g$
allows us
to model cases ranging from
little variation (top row in Figure \ref{fig:beta-g}) to
where a few sites have very different rates from the other sites (for
example, bottom right plot
in Figure~\ref{fig:beta-g}). In Table \ref{table:g-interpret} we
provide guidelines on the
interpretation of ${\log_{10}} g$.

%
%
\begin{table}[b]\vspace*{-3pt}
\tablewidth=138pt
\caption{Guidelines on the interpretation of the scaled variance $g$ on
the ${\log_{10}}$ scale}
\label{table:g-interpret}
\begin{tabular*}{138pt}{@{\extracolsep{4in minus 4in}}l@{\hspace*{20pt}}l@{}}
\hline
$\bolds{{\log_{10}} g}$ & \textbf{Variability} \\
\hline
$<\!\!-$3 & Very low \\
$-$3 to $-2$ & Low \\
$-$2 to $-1$ & Medium \\
$>\!\!-$1 & High \\
\hline
\end{tabular*}
\end{table}

We fit the model using Markov chain Monte Carlo (MCMC) methods
[Supplementary Material Section 3 in \citep{Fuetal2009supp}].
To check the reliability of the output of these methods, we
applied the algorithm to many simulated data sets, and also
confirmed
that point estimates of parameters from simpler versions
of the model we present here agreed closely with maximum likelihood
estimates obtained from an
expectation-maximization (EM) algorithm. See \citep{Fu2008} for further
details.

\subsection{Origins of data}
We collected DNA methylation patterns from the \textit{FMR1} locus (see
Section \ref{sec:introduction} for associated disease) on the X chromosome in leukocytes
using hairpin-bisulfite PCR [experimental conditions as in \citeauthor{Lairdetal2004}
 (\citeyear{Lairdetal2004}) and \citeauthor{Mineretal2004} (\citeyear{Mineretal2004}); also briefly discussed
in the Supplementary Material Section~2 in \citeauthor{Fuetal2009supp} (\citeyear{Fuetal2009supp})]. Due to
cell--cell variation, double-stranded methylation patterns
were collected from multiple cells in each individual sampled. The
data analyzed here contain 169 double-stranded
methylation patterns, each from a single cell, from 6 independent
normal females (15--33 cells or patterns per individual) at~22~CpG sites
(chrX: 146800867-146801008)
in the promoter region of the \textit{FMR1} locus. Each female cell has
two X chromosomes: one is hypermethylated, hence
primarily inactive, and the other hypomethylated, and hence primarily
active. The
data presented here are from the hypermethylated \textit{FMR1} locus on
the inactive X chromosome in each cell sampled.
Although this data set may be considered to be limited
in size, the data are unusual in their double-strandedness compared to
the single-stranded methylation data commonly produced
from high-throughput technologies. A small subset of these data, which
contains 33 methylation patterns at 7 CpG sites,
was published in \citep{Genereuxetal2005}.

\section{Results}
\label{sec:results}

We applied our model to the \textit{FMR1} data described above. Since
these data were collected from the
primarily hypermethylated (hence inactivated) X chromosome in normal females,
the methylation density is high as expected:
$81.9\%$ of all CpG dyads are methylated on both strands, $6.4\%$ are methylated
on just one strand, and $11.7\%$ are unmethylated on both strands [see
Section \ref{sec:introduction} and
\citet{Genereuxetal2005} for previous analysis of a subset of the data].

Here we treat those double-stranded methylation patterns from the six
different individuals as independent
samples from a single, homogeneous population of methylation patterns.
This treatment is effectively
equivalent to assuming no variations in $m_j$, $\mu_j$, ${\delta
_p}_j$ and ${\delta_d}
_j$ across the individuals. This
seems to be a reasonable starting point, given the current absence of
evidence for notable variations in at
least some of these parameters among individuals [\citet
{Stogeretal1997}]. In a more elaborate analysis,
however, we could relax this assumption by incorporating variability
across individuals into our hierarchical
model. Furthermore, our model does not distinguish between methylation
patterns from X chromosomes
inherited from the mother and those inherited from the father.
Information on the parental origins of a given
methylation pattern is not available for our \textit{FMR1} data.

The failure of bisulfite conversion rate, $b$, is relatively
straightforward to estimate directly for the
methylation patterns analyzed [Supplementary Material Section 2 in
\citet{Fuetal2009supp}].
We estimated $b$ to be 0.003 for our \textit{FMR1} data [\citet
{Lairdetal2004}] and in the analysis here
fixed it to be constant across sites. By comparison, the inappropriate
conversion rate $c$ is harder
to obtain directly. We estimate this rate in our data analysis.

We performed three independent runs of our MCMC fitting procedure, each
from a different starting point:
two runs of 1.44 million iterations, and a third run of~2.88 M
iterations (total compute time $\sim$ 160 hours
on a 2.4 GHz CPU). We sampled each MCMC run every 2 K iterations (or 4 K
for the third run) after discarding
the initial 20$\%$ of each run as burn-in. Trace plots displaying MCMC
samples versus iterations (not shown)
provided no indication of poor mixing. Histograms of key parameters
from different runs (not shown) also
agreed closely with one another. Results below come from pooling the
samples from the three runs. These
long runs were carried out to ensure convergence and may have exceeded
necessity; in fact, we achieved
similar results from much shorter pilot runs of 50 K iterations. When
using credible intervals to summarize
posterior distributions, we provide 80$\%$ coverage, which is not
unduly influenced by long tails of the
distributions.

\subsection{Rate of measurement error due to inappropriate bisulfite
conversion, and its variability}

\begin{figure}[t]

\includegraphics{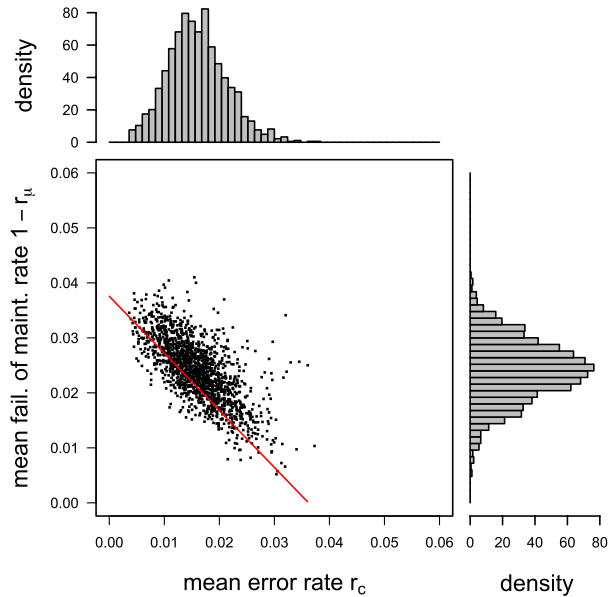}

\caption{Posterior distributions and scatter plot of mean failure of
maintenance rate, $1-r_\mu$, and
mean error rate, $r_c$, under the multi-site model for the \textit{FMR1}
data. The red line, $1-\mu=1.04 c + 0.04$,
indicates a predicted relationship for these estimates under a much
simpler analysis (see Section \protect\ref{sec:results.failure.rate}).}
\label{fig:hist.r.mu.c.1-22}\vspace*{-2pt}
\end{figure}

\begin{figure}

\includegraphics{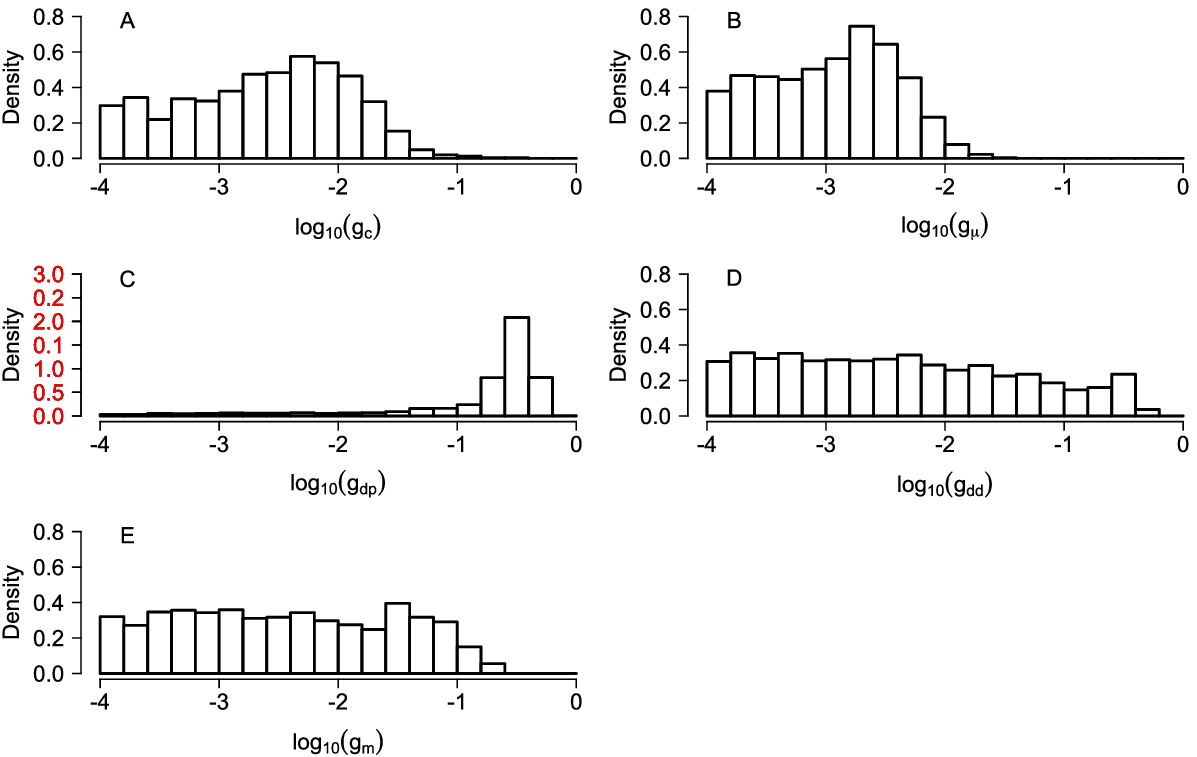}

\caption{Posterior distributions of $\log_{10}g$ for the \textit{FMR1}
data. $g$ in \textup{({A})--({D})} is the scaled
variance in the beta distribution assumed for \textup{({A})} measurement
error rate $c$ (due to inappropriate bisulfite
conversion); \textup{({B})} failure of maintenance rate $1-\mu$; \textup{({C})}
parent de novo rate ${\delta_p};$
and \textup{({D})} daughter de novo rate ${\delta_d}$. In \textup{({E})}, $g_m$ reflects
deviation
from the stationarity assumption. See Table \protect\ref{table:g-interpret} for
guidelines on the interpretation of values of ${\log_{10}} g$.
The $y$-axis in \textup{({C})} has a wide range.}
\label{fig:hist.g.1-22}
\end{figure}

The \textit{FMR1} data provide strong evidence for the occurrence of
inappropriate conversion error:
the posterior distribution for the mean error rate $r_c$ across CpG
sites is centered on 0.016, with 80$\%$ credible interval
(CI) of (0.009, 0.023) (top histogram in Figure \ref
{fig:hist.r.mu.c.1-22}), and there is little probability mass very near
0. The
posterior distribution for scaled variance $g_c$ is concentrated
on small values, suggesting that the error rate $c$ does not vary
greatly across CpG sites (Figure \ref{fig:hist.g.1-22}A),
in accord with experimental findings [\citet{Genereuxetal2008}].

\subsection{Failure of maintenance rate and its variability}
\label{sec:results.failure.rate}

We estimate the mean failure of maintenance rate $1-r_{\mu}$ across CpG
sites to be 0.024 (80$\%$ CI: 0.017--0.031; side histogram in
Figure \ref{fig:hist.r.mu.c.1-22}). MCMC samples of $1-r_\mu$ and $r_c$
show a striking linear relationship (Figure
\ref{fig:hist.r.mu.c.1-22}), suggesting a degree of unidentifiability
in these parameters. This relationship turns out to conform
very closely to predictions under a much simpler analysis based on
summarizing the \textit{FMR1} data by the overall proportions of methylated,
hemimethylated and unmethylated sites
$(p_M,p_H,p_U)=(0.82,0.064,0.116)$ [red line in Figure \ref
{fig:hist.r.mu.c.1-22};
see the  analysis in the Supplementary Material Section 4.1 in
\citet{Fuetal2009supp}]. This agreement between two very different
analysis approaches
suggests the robustness of the inference that $1-r_\mu$ and $r_c$ lie
close to this line. The fact that our MCMC samples
are concentrated on only part of this line reflects the additional
information we are able
to extract from the full data by making more detailed modeling
assumptions as stated in Section \ref{sec:methods.basic}.
The inference, of course, must then be less robust to deviations from
these assumptions.

Regarding variability of $1-\mu$ across CpG sites, the data suggest
that this variability is low, since the posterior for $g_\mu$ is
concentrated around small values (Figure~\ref{fig:hist.g.1-22}B).

\subsection{De novo methylation rates and their variability}
\label{sec:results.denovo}

\begin{figure}[t]

\includegraphics{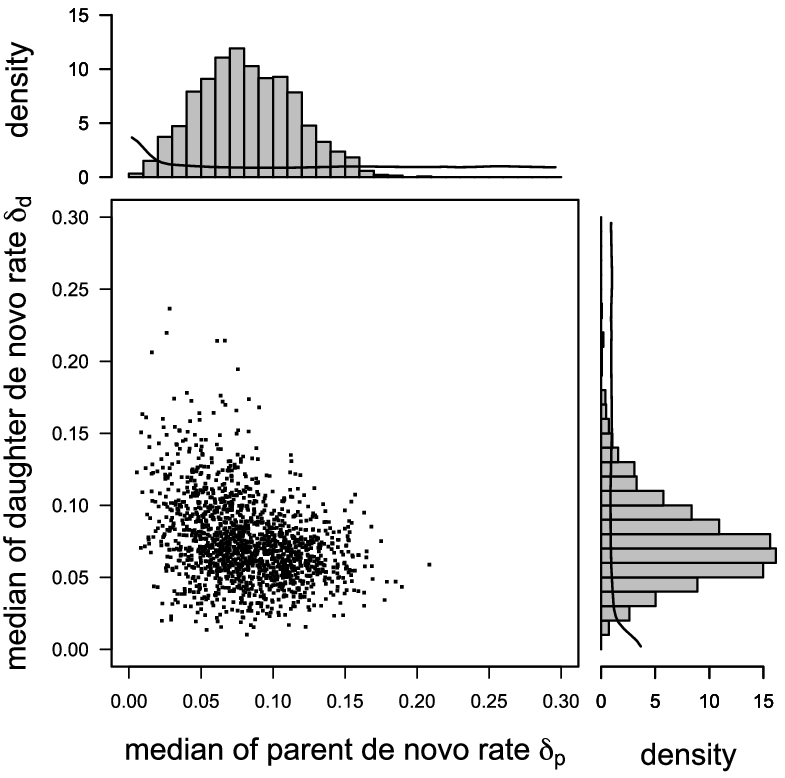}

\caption{Posterior distributions and scatter plot of median de novo
rates ${\delta_p}$ and ${\delta_d}$
under the multi-site model for the \textit{FMR1} data. Smooth curves in
the histograms are density functions of the median of
22 beta random variables, each corresponding to a parent (or daughter)
de novo rate at a CpG site. The two density curves
are identical because the prior distributions for the rates are identical.}
\label{fig:hist.dpdd.1-22}
\end{figure}

Our results suggest that de novo rates may be substantially larger than
failure of maintenance rates (which can happen even at stationarity):
the posterior distribution
for the median parent  and daughter de novo rates are centered on 0.08
and 0.07, respectively, with very low
probability mass near 0 (histograms in Figure \ref{fig:hist.dpdd.1-22}; compare with Figure \ref{fig:hist.r.mu.c.1-22}).
These high rate estimates are consistent with, and may partly explain,
the high overall methylation rates in this genomic region.
There is, however, considerable uncertainty in these estimates:
80\% CIs are 0.04--0.13 and 0.04--0.11, respectively (histograms in
Figure \ref{fig:hist.dpdd.1-22}).
Note that the scatter plot shows that these two parameters are not
independent, a posteriori: in particular, it is unlikely
that both de novo rates are at the upper end of these CIs (no MCMC
sample in the scatter plot in Figure \ref{fig:hist.dpdd.1-22}
has both rates $>$ 0.13).

\begin{figure}[bt]

\includegraphics{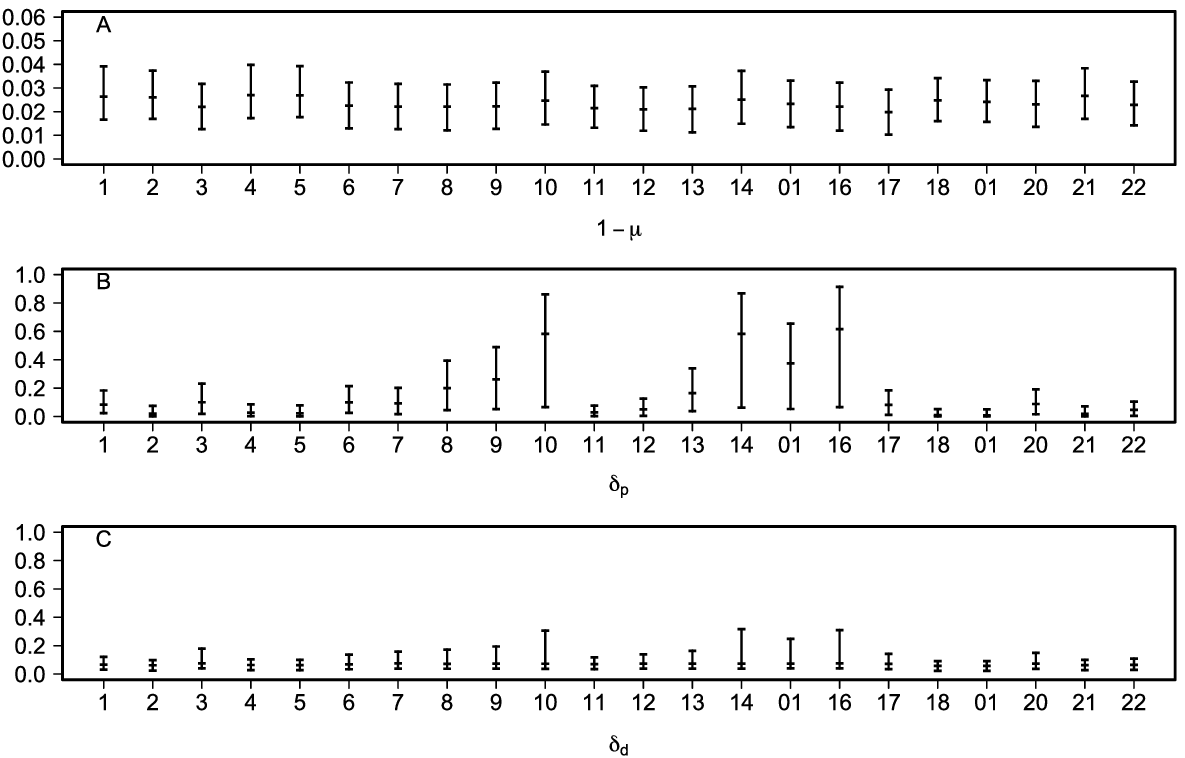}

\caption{Median and 80$\%$ credible intervals of CpG-site-specific
estimates of \textup{({A})} failure of maintenance rates $1-\mu$,
\textup{({B})} parent de novo rates ${\delta_p}$ and \textup{({C})} daughter de novo
rates ${\delta_d}$
under the multi-site model for the \textit{FMR1} data. The numbering of
the CpG sites follows the convention established in \protect\citet{Stogeretal1997}.}
\label{fig:boxplots.rates.1-22}
\end{figure}

One biological question of interest is whether or not de novo events
occur on both parent and daughter strands. We do not
conduct a formal test of hypotheses here, but
we note that the posterior
distribution of the median of each de novo rate has little probability
mass near 0 (Figure \ref{fig:hist.dpdd.1-22}), in contrast to the prior
distribution,
providing informal support for both parent and daughter de novo events
occurring.

Regarding variability across sites, the data are uninformative for
variability in the daughter de novo rate: the posterior for ${\log
_{10}} g_{dd}$
is flat over the whole support (Figure \ref{fig:hist.g.1-22}D). In
contrast, the parent de novo rate ${\delta_p}$ may vary considerably
across sites:
${\log_{10}} g_{dp}$ is concentrated on large values (see Figure \ref
{fig:hist.g.1-22}C and compare with the bottom right panel in
Figure \ref{fig:beta-g}). Furthermore, a few outlying sites have
possibly high rates (Figure \ref{fig:boxplots.rates.1-22}B, in contrast
to little variation in
$1-\mu$ in Figure \ref{fig:boxplots.rates.1-22}A and in ${\delta_d}$ in
Figure \ref{fig:boxplots.rates.1-22}C), which may have a strong influence on the
mean value
across sites. This large variability makes it difficult to estimate
mean de novo rates and renders them misleading in summarizing
site-specific ${\delta_p}$s. Therefore, we have chosen to report the
median de
novo rates.

The observation that ${\delta_p}$ may vary considerably across sites brings
into question the suitability of our assumption of
a single beta distribution for these rates, since this assumption
has limited flexibility in dealing with potential outliers. To examine
this issue, we modified our model to
allow the de novo rate parameters to follow a mixture of two beta
distributions, where the component
corresponding to the outlying sites was assumed to be $\operatorname{Uniform}(0,1)$
[that is, $\operatorname{Beta}(\alpha=1,\beta=1)$]. Analyses using this model continued
to suggest that some sites (specifically sites~10, 14, 15 and 16) may
have substantially
higher parent de novo rates than others (see \mbox{\citet{Fu2008}} for
further details).
Indeed, the data at these four sites are characterized by
particularly small numbers of unmethylated CpG dyads (0 at site 16, 1
at sites~10 and 14, and 3 at site 15, in contrast to
the median of 20 at other sites).

\begin{figure}[b]

\includegraphics{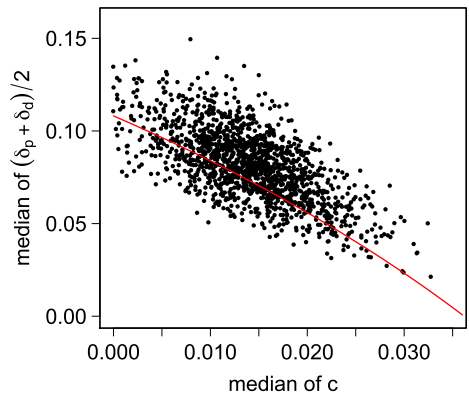}

\caption{Joint posterior distribution of the median of the average of
the parent and daughter de novo rates, $({\delta_p}+{\delta_d})/2$,
and the median
of the error rate $c$ for the \textit{FMR1} data. The red curve, $({\delta
_p}+{\delta_d}
)/2 = 0.44 + 0.05/(c-0.15)$,
indicates a predicted relationship for these estimates under a much
simpler analysis (see text).}
\label{fig:scatter.dpdd.1-22}
\end{figure}

Our analysis differs from previous analyses by accounting for
measurement errors which have rate $c$. To gain insight into
how incorporating error rates affects estimated de novo rates, we
examined the joint posterior distribution of $c$ and
the average de novo rate (Figure \ref{fig:scatter.dpdd.1-22}). As in
the analagous plot for failure of maintenance rate (Figure \ref
{fig:hist.r.mu.c.1-22}),
posterior samples here also lie close to a line, which is in close
agreement with a simple analysis based on summary statistics
[Supplementary Material Section 4.1 in {\citet{Fuetal2009supp}}].
Remarks made above in Section \ref{sec:results.failure.rate} regarding
robustness of the conclusions apply equally here.
%

Another important novel contribution of our analysis is that, by
modeling the strand information in multi-site data, we can
distinguish, at least in principle, between the two different types of
de novo events. This novel feature makes it possible to
draw several conclusions mentioned above, particularly that the data
support the occurrence of both parent and daughter
de novo events, and that the data provide different information on the
variability of ${\delta_p}$ and ${\delta_d}$.
However, due to the relative complexity of our model, it is difficult
to identify the source of the information that distinguishes
daughter de novo events from parent de novo events. To gain insight, we
examined in some detail the multi-site likelihood for a single
methylation pattern  [Supplementary Material Section 4.2
in \citet{Fuetal2009supp}]. A~conclusion from
this investigation is that, assuming stationarity (or, in fact, under
weaker assumptions), data on methylation patterns where one strand is
much more
methylated than the other will tend to favor large estimates of
${\delta_p}$
relative to ${\delta_d}$. Additionally, the more methylated
strand will tend to be the parent strand. Thus, an intuitive
explanation of our inference that sites 10, 14, 15 and 16 have
large ${\delta_p}$ is that some patterns, with large differences in the
methylation density on the two strands, are hemimethylated
at these sites (with the methylated CpG more likely to be on the
overall more methylated strand). The novel insights into
the de novo rates we gain here are further discussed in Section \ref
{sec:conclusions}.


\subsection{Stationarity}

To examine the extent to which the data are consistent with a
stationary model, we consider the
posterior distribution of $\log_{10} g_m$, which reflects deviations
from stationarity (Figure \ref{fig:hist.g.1-22}E).
This posterior largely follows the uniform prior,
except that large values are excluded. We conclude
that the data do not exhibit large deviations from stationarity,
although they do not
provide strong support for the strict stationarity assumption either.

\subsection{Impact of bisulfite conversion errors on the estimation of
failure of maintenance and de novo
methylation rates}
\label{sec:error.impact}
Our analyses above incorporate both types of bisulfite conversion
errors, which have not
been accounted for in previous analyses of methylation patterns [see,
e.g.,
\citet{Genereuxetal2005}; \citet{Lairdetal2004};
\citet{Ushijimaetal2003};
\citet{Pfeiferetal1990}].
It seems possible that our incorporation of measurement error may be
the main reason for
discrepancies between our estimates of rates of methylation events and
estimates from
these previous analyses. To assess this, we reran the multi-site model
on the \textit{FMR1} data,
setting the two measurement error rates $b$ and $c$ to be 0, which corresponds
to ignoring both types of bisulfite conversion errors. We
carried out three independent runs from different starting points. Each
run consisted of 1.44 million iterations, including~20$\%$ burn-in, and took about 38 hours. These runs gave consistent
results, so we pooled the three runs to
produce the posterior distributions.

\begin{figure}

\includegraphics{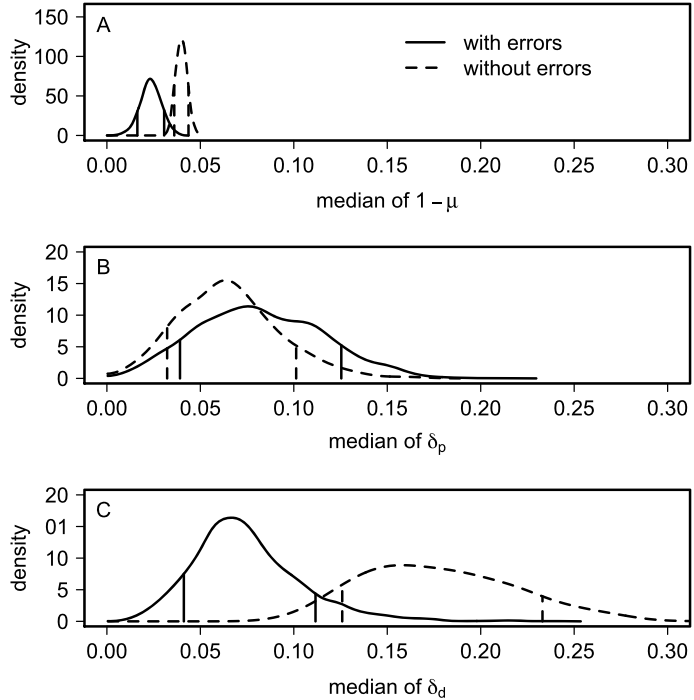}

\caption{Impact of measurement errors (due mainly to inappropriate
bisulfite conversion error) on the inference of the rates of
methylation events.
Solid lines incorporate these errors, whereas dashed lines do not. From
top to bottom are posterior densities of
medians of \textup{({A})} failure of maintenance rate $1-\mu$, \textup{({B})}
parent de novo rate ${\delta_p}$ and \textup{({C})} daughter de novo rate
${\delta_d}$.
Vertical bars indicate the~80$\%$ credible interval (10$\%$ and 90$\%$
percentiles) for each density.}
\label{fig:density.med.1-22}
\end{figure}

Our results show that incorporating measurement errors indeed has
substantial effects on the inference of failure of maintenance
rate $1-\mu$ and daughter de novo rate ${\delta_d}$ (Figure \ref
{fig:density.med.1-22}A and C) but little effects on parent de novo rate
${\delta_p}$ (Figure \ref{fig:density.med.1-22}B). Estimates of these two
rates under the no-error model are largely consistent
with previous results [\citet{Lairdetal2004}; \citet{Genereuxetal2005}]. This
comparison suggests that whether or not measurement
error is accounted for may have been an important factor in producing
different inferences.

\section{Discussion and conclusions}
\label{sec:conclusions}
We have developed a statistical model for double-stranded DNA
methylation patterns to
investigate a central problem in epigenetic biology:
the transmission fidelity of DNA methylation patterns in somatic
mammalian cells.
Our modeling approach addresses several challenges that are inherent in
these data and that have
not been approached by previous methods. Key innovations of our model
include the incorporation
of measurement error and the incorporation of available ``phase''
information;
that is, which hemimethylated CpG dyads are methylated on the same
strand, by examining multiple
sites simultaneously. The first innovation is important because, as we
have shown, measurement
error has a substantial effect on estimates of fidelity rates. The second
is important because it allows us both to separately estimate parent
and daughter de novo rates,
and to relax the strict stationarity assumption that underlies most
existing approaches
[see, for example, \citet{OttoWalbot1990}; \citet{Pfeiferetal1990};
\citet{Genereuxetal2005}].

By applying our new model to the \textit{FMR1} data, we gained several new
insights into
methylation transmission fidelity rates. Below we
summarize our findings and compare them with other studies:
\begin{enumerate}
\item Inappropriate bisulfite conversion can be a significant source of
measurement error. We estimated the
mean rate of this error in our data set to be 0.016 (80$\%$ CI:
0.009--0.023). As far as we are aware, ours is
the first estimate of this inappropriate conversion rate obtained from
genomic methylation pattern data
that are double-stranded and molecularly-validated (see Section \ref
{sec:introduction}
for detail on data collection). Our estimate of this rate is lower than
that obtained by \citet{Genereuxetal2008}
using synthetic oligonucleotides (average rate 0.035; 95$\%$ confidence
interval: 0.027--0.049). This difference may derive, in
part, from the different lengths of the DNAs used in the two
experiments [\citet{Genereuxetal2008}].
\item We estimated the mean maintenance rate $\mu$ to be 0.976
(80$\%$ CI: 0.969--0.983). This is higher than previous estimates for
similar data
[\citet{Lairdetal2004}; \citet{Genereuxetal2005}],
which can be mostly explained by the fact that these previous analyses
did not account for bisulfite conversion errors.
On the other hand, \citep{Pfeiferetal1990} estimated the maintenance
rate to be much higher
(about 0.999), which is mainly due to a high overall methylation
density ($\sim$0.98) at the site analyzed.
\item We found suggestive evidence that de novo events occur on both
parent and daughter strands,
in that posterior distributions for both parent and daughter de novo
rates have little
probability mass near 0. Previous empirical studies have asked whether
de novo events can occur on the
parent strand [\citet{Kappler1970}; \citet{Adams1971};
\citet{SchneidermanBillen1973};
\citet{Bird1978}], yielding conflicting
conclusions for different cell types. Recent analyses still could not address
this question because phase information was either not available
[\citet{Pfeiferetal1990}; \citet{Ushijimaetal2003}] or not incorporated in
their models [\citet{Lairdetal2004};\break \citet{Genereuxetal2005}].
To accommodate those limitations, \citet{Pfeiferetal1990} estimated
the total de novo rate as a whole, whereas
\citet{Lairdetal2004} and \citet{Genereuxetal2005} imposed additional
constraints that are equivalent to estimating
the total de novo rates. Potential implications of a positive parent de novo
rate are discussed in \citet{Genereux2009}.
\item We also found some evidence that parent de novo rates vary
considerably across sites.
In particular, sites 10, 14, 15 and 16 in our data may experience
unusually high parent de novo rates.
Analyses of the data at these sites individually using the single-site
approach from \citet{Genereuxetal2005}
also suggested potentially large values for the total de novo rate at
these sites, although
the single-site approach was unable to separately estimate the de novo
rates on parent and daughter strands.
\end{enumerate}

Some previous studies estimated an overall methylation transmission fidelity
rate, tracking methylation patterns over one [\citet{Bird1978}] or more
[\citet{Ushijimaetal2003}] rounds of DNA replication.
Different experimental techniques and sampling procedures used in these
studies led to data of very different types
from that of our \textit{FMR1} data. A fair comparison of the results is
difficult to carry out
because of these differences and is therefore not addressed here.

A limitation of our model is the assumption that methylation events
occur independently across CpG sites. This
assumption does not seem to hold in practice, especially for
maintenance events, in light of current research on
methylation enzymes [\citet{Vilkaitisetal2005}; \citet{Goyaletal2006}]. It is
therefore of great interest to study the
dependence structure. In separate work we developed statistical models
to incorporate the dependence
[\citet{Fu2008}]. Our preliminary results there yielded similar estimates
of at least the mean rates (parameter $r$)
of the methylation events to the estimates in this paper.

As more hairpin-bisulfite PCR data become available, the new
statistical analysis methods described here may continue to
provide novel biological insights into epigenetic fidelity. The
estimation precision will improve
as new experimental protocols yield data with lower measurement error
rates [\citet{Genereuxetal2008}]
and lead to better estimates of de novo methylation rates. With our
statistical methods one can investigate differences
among fidelity rates in different genomic regions. For example, our
model can be applied also to sparsely methylated
CpG islands where de novo rates may take on a wider range of values
than in densely methylated regions
[\citet{Ushijimaetal2003}; \citet{Lairdetal2004}]. Furthermore, relaxation of
the stationarity condition gives our methods
great flexibility to examine the transmission of methylation patterns
in cases where methylation densities are
dynamic rather than stationary. Many of these cases have important
clinical and pharmaceutical implications;
they include early developmental stages characterized by loss and
re-establishment of methylation patterns [\citet{Reiketal2001}], aging
during which
methylation patterns may change over time in at least certain cell
types [\citet{WilsonJones1983}], and in several types of
cancer in which methylation patterns change rapidly over cell
generations [\citet{Fosteretal1998}].
These cases will pose new challenges. For instance, sets of methylation
patterns collected
from cancer patients may be sampled from a mixture of cancer cells and
normal cells. Successful analysis
of such data must account for the existence of these subpopulations, a
challenging yet
intriguing research direction for the future.


\section*{Acknowledgments}
The authors would like to acknowledge the invaluable contribution
Brooks Miner made to the collection
of the \textit{FMR1} data analyzed in this paper, a small subset of which
was presented and analyzed
previously by \citet{Genereuxetal2005}. Thanks also go to
Krista Gile, Peter Hoff, Vladmir Minin and Elizabeth Thompson for stimulating
discussions and thought-provoking questions. The authors are grateful
to the editor and two anonymous referees for
their excellent comments and questions, which have greatly improved
this paper.

\begin{supplement}[id=suppA]
\sname{Supplement A}
\stitle{Appendices}
\slink[doi]{10.1214/09-AOAS297SUPPA}
\slink[url]{http://lib.stat.cmu.edu/aoas/297/supplement_A.pdf}
\sdatatype{.pdf}
\sdescription{The pdf file contains biological background,
experimental design issues, Markov chain Monte Carlo (MCMC)
procedures and likelihood analyses for special cases.}
\end{supplement}
\begin{supplement}[id=suppB]
\sname{Supplement B}
\stitle{Data and MCMC code}
\slink[doi]{10.1214/09-AOAS297SUPPB}
\slink[url]{http://lib.stat.cmu.edu/aoas/297/supplement_B.zip}
\sdatatype{.zip}
\sdescription{The zip file contains the \textit{FMR1} data analyzed in
this paper, the
R code that implements the MCMC procedure and MCMC outputs summarized
and displayed in Section \ref{sec:results}.}
\end{supplement}

\printaddresses


\begin{thebibliography}{}

\bibitem[\protect\citeauthoryear{Adams}{1971}]{Adams1971}
\textsc{Adams, R. L.} (1971).
 Methylation of newly synthesized and older deoxyribonucleic acid.
\textit{Biochem.~J.} \textbf{123} 38.

\bibitem[\protect\citeauthoryear{Bird}{1978}]{Bird1978}
\textsc{Bird, A.} (1978).
 Use of restriction enzymes to study eukaryotic {DNA} methylation:
{II}. The symmetry of methylated sites supports semi-conservative
copying of
the methylation pattern.
\textit{J. Mol. Biol.} \textbf{118} 49--60.

\bibitem[\protect\citeauthoryear{Bird}{2002}]{Bird2002}
\textsc{Bird, A.} (2002).
{DNA} methylation patterns and epigenetic memory.
\textit{Genes Dev.} \textbf{16} 6--21.

\bibitem[\protect\citeauthoryear{Bird}{2007}]{Bird2007}
\textsc{Bird, A.} (2007).
 Perceptions of epigenetics.
\textit{Nature} \textbf{447} 396--398.

\bibitem[\protect\citeauthoryear{Chen and Riggs}{2005}]{ChenRiggs2005}
\textsc{Chen, Z.} and \textsc{Riggs, A. D.} (2005).
 Maintenance and regulation of {DNA} methylation patterns in mammals.
\textit{Biochem. Cell Biol.} \textbf{83} 438--448.

\bibitem[\protect\citeauthoryear{Ehrlich et al.}{1982}]{Ehrlichetal1982}
\textsc{Ehrlich, M., Gama-Sosa, M. A., Huang, L. H., Midgett, R. M., Kuo, K. C., McCune,~R.~A.}
and \textsc{Gehrke, C.} (1982).
 Amount and distribution of 5-methylcytosine in human {DNA} from
different types of tissues or cells.
\textit{Nucleic Acids Res.} \textbf{10} 2709--2721.

\bibitem[\protect\citeauthoryear{Foster et al.}{1998}]{Fosteretal1998}
\textsc{Foster, S. A., Wong, D. J., Barrett, M. T.} and \textsc{Galloway, D. A.} (1998).
 Inactivation of p16 in human mammary epithelial cells by {C}p{G}
island methylation.
\textit{Mol. Cell. Biol.} \textbf{18} 1793--1801.

\bibitem[\protect\citeauthoryear{Fu et al.}{2009}]{Fuetal2009supp}
\textsc{Fu, A., Genereux, D.,  St{\"{o}}ger, R., Laird, C.} and \textsc{Stephens, M.}
(2009).
 Supplement to ``{S}tatistical inference of transmission
fidelity of
{DNA} methylation patterns over somatic cell divisions in mammals.''
DOI: \href{http://dx.doi.org/10.1214/09-AOAS297SUPPA}{10.1214/09-AOAS297SUPPA},
DOI:
\href{http://dx.doi.org/10.1214/09-AOAS297SUPPB}{10.1214/09-AOAS297SUPPB}.

\bibitem[\protect\citeauthoryear{Fu}{2008}]{Fu2008}
\textsc{Fu, Q.} (2008).
Models and inference of transmission of {DNA} methylation
patterns in mammalian somatic cells.
 Ph.D. dissertation, Univ. Washington.
\MR{2717564}

\bibitem[\protect\citeauthoryear{Genereux}{2009}]{Genereux2009}
\textsc{Genereux, D. P.} (2009).
 Asymmetric strand segregation: Epigenetic costs of genetic fidelity?
\textit{PLoS Genet.} \textbf{5} e1000509.
 DOI: \href{http://dx.doi.org/10.1371/journal.pgen.1000509}{10.1371/journal.pgen.1000509}.

\bibitem[\protect\citeauthoryear{Genereux et al.}{2008}]{Genereuxetal2008}
\textsc{Genereux, D. P., Johnson, W. C., Burden, A. F., S{t\"{o}}ger, R.} and
\textsc{Laird, C. D.} (2008).
 Errors in the bisulfite conversion of {DNA}: Modulating
inappropriate- and failed-conversion frequencies.
\textit{Nucleic Acids Res.} \textbf{36} e150.

\bibitem[\protect\citeauthoryear{Genereux et al.}{2005}]{Genereuxetal2005}
\textsc{Genereux, D. P., Miner, B. E., Bergstrom, C. T.} and \textsc{Laird, C. D.} (2005).
 A population-epigenetic model to infer site-specific methylation
rates from double-stranded {DNA} methylation patterns.
\textit{Proc. Natl. Acad. Sci. USA} \textbf{102} 5802--5807.

\bibitem[\protect\citeauthoryear{Goyal, Reinhardt and Jeltsch}{2006}]{Goyaletal2006}
\textsc{Goyal, R., Reinhardt, R.} and \textsc{Jeltsch, A.} (2006).
 Accuracy of {DNA} methylation pattern perservation by the {D}nmt1
methyltransferase.
\textit{Nucleic Acids Res.} \textbf{34} 1182--1188.

\bibitem[\protect\citeauthoryear{Jones and Baylin}{2002}]{JonesBaylin2002}
\textsc{Jones, P. A.} and \textsc{Baylin, S. B.} (2002).
 The fundamental role of epigenetic events in cancer.
\textit{Nature Rev. Genet.} \textbf{3} 415--428.

\bibitem[\protect\citeauthoryear{Kangaspeska et al.}{2008}]{Kangaspeskaetal2008}
\textsc{Kangaspeska, S., Stride, B., {M}\'{e}tivier, R., Polycarpou-Schwarz,
M., Ibberson, D., Carmouche, R. P., Benes, V., Gannon, F.} and \textsc{Reid, G.} (2008).
 Transient cyclical methylation of promoter {DNA}.
\textit{Nature} \textbf{452} 112--116.

\bibitem[\protect\citeauthoryear{Kappler}{1970}]{Kappler1970}
\textsc{Kappler, J. W.} (1970).
 The kinetics of {DNA} methylation in cultures of a mouse
adrenal cell
line.
\textit{J. Cell. Physiol.} \textbf{75} 21--31.

\bibitem[\protect\citeauthoryear{Laird}{1987}]{Laird1987}
\textsc{Laird, C. D.} (1987).
 Proposed mechanism of inheritance and expression of the human
fragile-{X}~syndrome of mental retardation.
\textit{Genetics} \textbf{117} 587--599.

\bibitem[\protect\citeauthoryear{Laird et al.}{2004}]{Lairdetal2004}
\textsc{Laird, C. D., Pleasant, N. D., Clark, A. D., Sneeden, J. L. S., Hassan, K. M.
A., Manley, N. C., Vary, J. C., Morgan, T., Hansen, R. S.} and \textsc{St{\"{o}}ger, R.}
(2004).
 Hairpin-bisulfite {PCR}: Assessing epigenetic methylation
patterns on
complementary strands of individual {DNA} molecules.
\textit{Proc. Natl. Acad. Sci. USA} \textbf{101} 204--209.

\bibitem[\protect\citeauthoryear{Laird}{2003}]{Laird2003}
\textsc{Laird, P. W.} (2003).
 The power and the promise of {DNA} methylation markers.
\textit{Nature Rev. Cancer} \textbf{3} 253--266.

\bibitem[\protect\citeauthoryear{M{\'{e}}tivier et al.}{2008}]{Metivieretal2008}
\textsc{M{\'{e}}tivier, R., Gallais, R., Tiffoche, C., Le P{\'{e}}ron, C., Jurkowska, R.
Z., Carmouche,~R.~P., Ibberson, D., Barath, P., Demay, F., Reid, G.,
Benes, V., Jeltsch,~A., Gannon, F.} and \textsc{Salbert, G.} (2008).
 Cyclical {DNA} methylation of a transcriptionally active promoter.
\textit{Nature} \textbf{452} 45--52.

\bibitem[\protect\citeauthoryear{Miner et al.}{2004}]{Mineretal2004}
\textsc{Miner, B. E., St{\"{o}}ger, R. J., Burden, A. F., Laird, C. D.} and \textsc{Hansen, R. S.}
(2004).
 Molecular barcodes detect redundancy and contamination in
hairpin-bisulfite {PCR}.
\textit{Nucleic Acids Res.} \textbf{32} e135.

\bibitem[\protect\citeauthoryear{Otto and Walbot}{1990}]{OttoWalbot1990}
\textsc{Otto, S.} and \textsc{Walbot, V.} (1990).
{DNA} methylation in eukaryotes: Kinetics of demethylation
and de
novo methylation during the life cycle.
\textit{Genetics} \textbf{124} 429--437.

\bibitem[\protect\citeauthoryear{Pfeifer et al.}{1990}]{Pfeiferetal1990}
\textsc{Pfeifer, G., Steigerwald, S., Hansen, R., Gartler, S.} and \textsc{Riggs, A.} (1990).
 Polymerase chain reaction-aided genomic sequencing of an {X}
chromosome-linked {C}p{G} island: Methylation patterns suggest clonal
inheritance, {C}p{G} site autonomy, and an explanation of activity state
stability.
\textit{Proc. Natl. Acad. Sci. USA} \textbf{87} 8252--8256.

\bibitem[\protect\citeauthoryear{Reik, Dean and Walter}{2001}]{Reiketal2001}
\textsc{Reik, W., Dean, W.} and \textsc{Walter, J.} (2001).
 Epigenetic reprogramming in mammalian development.
\textit{Science} \textbf{293} 1089--1093.

\bibitem[\protect\citeauthoryear{Robertson and Wolffe}{2000}]{RobertsonWolffe2000}
\textsc{Robertson, K. D.} and \textsc{Wolffe, A. P.} (2000).
{DNA} methylation in health and disease.
\textit{Nature Rev. Genet.} \textbf{1} 11--19.

\bibitem[\protect\citeauthoryear{Schneiderman and Billen}{1973}]{SchneidermanBillen1973}
\textsc{Schneiderman, M. H.} and \textsc{Billen, D.} (1973).
 Methylation rapidly reannealing {DNA} during the cell cycle of
chinese hamster cells.
\textit{Biochim. Biophys. Acta} \textbf{308} 352--360.

\bibitem[\protect\citeauthoryear{St{\"{o}}ger et al.}{1997}]{Stogeretal1997}
\textsc{St{\"{o}}ger, R., Kajimura, T. M., Brown, W. T.} and \textsc{Laird, C. D.} (1997).
 Epigenetic variation illustrated by {DNA} methylation
patterns of the
fragile-{X} gene {\textit{FMR1}}.
\textit{Hum. Mol. Genet.} \textbf{6} \mbox{1791--1801}.

\bibitem[\protect\citeauthoryear{Ushijima et al.}{2003}]{Ushijimaetal2003}
\textsc{Ushijima, T., Watanabe, N., Okochi, E., Kaneda, A., Sugimura, T.} and \textsc{Miyamoto, K.}
(2003).
 Fidelity of the methylation pattern and its variation in the genome.
\textit{Genome Res.} \textbf{13} \mbox{868--874}.

\bibitem[\protect\citeauthoryear{Vilkaitis et al.}{2005}]{Vilkaitisetal2005}
\textsc{Vilkaitis, G., Suetake, I., Klima{\v{s}}auskas, S.} and \textsc{Tajima, S.} (2005).
 Processive methylation of hemimethylated {C}p{G} sites by mouse
{D}nmt1 {DNA} methyltransferase.
\textit{J. Biol. Chem.} \textbf{280} \mbox{64--72}.

\bibitem[\protect\citeauthoryear{Wilson and Jones}{1983}]{WilsonJones1983}
\textsc{Wilson, V. L.} and \textsc{Jones, P. A.} (1983).
{DNA} methylation decreases in aging but not in immortal cells.
\textit{Science} \textbf{220} 1055--1057.

\end{thebibliography}
\end{document}